# Trustworthiness of Legal Considerations for the Use of LLMs in Education

Sara Alaswad
*College of Information Technology*
*Ahlia University*
Manama, Bahrain
salaswad@ahlia.edu.bh

Tatiana Kalganova
*Brunel University of London*
London, UK
tatiana.kalganova@brunel.ac.uk

Wasan Awad
*College of Information Technology*
*Ahlia University*
Manama, Bahrain
wawad@ahlia.edu.bh

***Abstract*— As Artificial Intelligence (AI)—particularly Large Language Models (LLMs)—becomes increasingly embedded in education systems worldwide, ensuring their ethical, legal, and contextually appropriate deployment has become a critical policy concern. This paper offers a comparative analysis of AI-related regulatory and ethical frameworks across key global regions, including the European Union, United Kingdom, United States, China, and Gulf Cooperation Council (GCC) countries. It maps how core trustworthiness principles—such as transparency, fairness, accountability, data privacy, and human oversight—are embedded in regional legislation and AI governance structures. Special emphasis is placed on the evolving landscape in the GCC, where countries are rapidly advancing national AI strategies and education-sector innovation. To support this development, the paper introduces a Compliance-Centered AI Governance Framework tailored to the GCC context. This includes a tiered typology and institutional checklist designed to help regulators, educators, and developers align AI adoption with both international norms and local values. By synthesizing global best practices with region-specific challenges, the paper contributes practical guidance for building legally sound, ethically grounded, and culturally sensitive AI systems in education. These insights are intended to inform future regulatory harmonization and promote responsible AI integration across diverse educational environments.**

## I. Introduction

The rapid advancement of Large Language Models (LLMs) has transformed various sectors, including education. These models, powered by Artificial Intelligence (AI), offer significant potential for personalized learning, automated assessment, and academic support [1]. However, their integration into educational settings raises critical legal concerns, including data privacy, intellectual property rights, academic integrity, and bias mitigation [2].

Central to these discussions is the concept of trustworthiness in AI. Trustworthy AI refers to systems that operate transparently, fairly, reliably, and in alignment with both legal standards and ethical principles. Key dimensions include transparency, fairness and non-discrimination, reliability and safety, accountability, data privacy, and the necessity for human oversight [3]. In the context of education, these principles are vital to ensuring that LLMs provide accurate, unbiased support without enabling academic misconduct or compromising student data.

Although regulations such as the General Data Protection Regulation (GDPR) in Europe and the Family Educational Rights and Privacy Act (FERPA) in the United States offer foundational protections, their sufficiency and adaptability in the face of AI-specific challenges remain debatable [4]. Additionally, legal ambiguities concerning the ownership of AI-generated content and institutional liability necessitate further exploration [5].

This paper contributes a comparative legal analysis of AI governance in education across the EU, UK, USA, China, and GCC countries, focusing on trustworthiness principles such as transparency, fairness, and accountability. It fills a gap in existing literature by highlighting emerging AI frameworks in the GCC, offering visual mappings of global regulatory approaches, and providing practical insights for policymakers balancing innovation with legal and ethical obligations.

In response to the region's fast-paced development and emerging regulatory frameworks, the paper also proposes a Compliance-Centered AI Governance Framework tailored to the GCC. This includes a tiered compliance typology and a practical checklist that support alignment with international standards while respecting local cultural and legal contexts. Together, these contributions aim to guide educators, policymakers, and developers toward responsible, context-sensitive AI integration in education.

## II. Methodology

This study adopts a comparative legal analysis approach to examine the regulatory and ethical frameworks governing AI in education across selected jurisdictions — including the EU, UK, USA, China, and the GCC. The selection was based on three criteria: (1) presence of formal AI legislation or national strategy, (2) inclusion of data protection or AI ethics principles, and (3) relevance to educational applications, particularly LLMs.

Data were gathered from official legal documents, policy reports, and peer-reviewed academic literature published between 2016 and 2025. Each framework was coded and compared according to key trustworthiness dimensions—Transparency, Fairness and Non-Discrimination, Reliability and Safety, Accountability, Data Privacy and Human Oversight—adapted from OECD (2021) and UNESCO (2023) AI ethics guidelines. The analysis aimed to identify patterns of convergence and divergence in AI governance, especially regarding compliance requirements in educational contexts.

## III. Regulatory Framework

### A. *Global Restrictions on ChatGPT: Implications for Legal Trustworthiness*

Several countries have implemented bans on ChatGPT for varying durations and reasons, reflecting diverse regulatory and political approaches to AI deployment. A global overview of these restrictions is presented in Fig. 1, which categorizes countries by the duration and rationale of their bans—ranging from privacy concerns to broader political restrictions. For instance, Italy's temporary ban (31 days) due to privacy issues

This work has been funded by the European Union. Views and opinions expressed are however those of the authors only and do not necessarily reflect those of the European Union or European Commission-EU. Neither the European Union nor the granting authority can be held responsible for them.

demonstrates agile regulatory oversight, whereas persistent bans in countries like North Korea and Syria reflect enduring digital control policies [6], [7].

These varied responses underscore the geopolitical and legal complexities surrounding AI governance. Understanding these dynamics is vital for evaluating the trustworthiness of legal frameworks in education, as similar tensions between innovation and regulation will likely shape institutional decisions on AI integration.

While ChatGPT has been the focal point of regulatory responses, the growing variety of LLMs—including those developed by USA, Chinese, and European entities—raises questions about how the origin of a model influences its perceived trustworthiness and regulatory acceptance, especially in education. This geopolitical dimension adds another layer to AI governance that goes beyond technical capabilities.

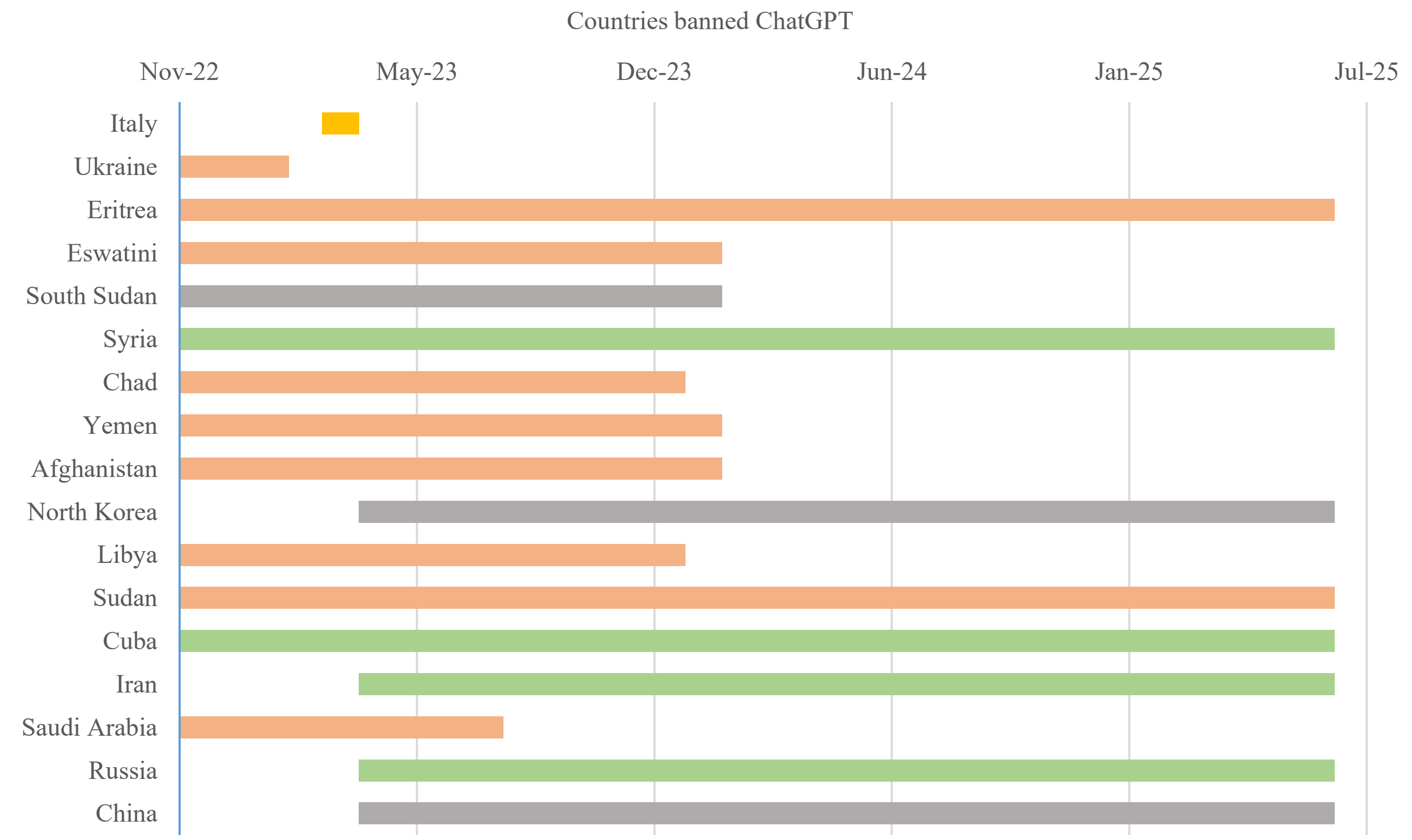

| | China | Russia | Saudi Arabia | Iran | Cuba | Sudan | Libya | North Korea | Afghanistan | Yemen | Chad | Syria | South Sudan | Eswatini | Eritrea | Ukraine | Italy |
|---|---|---|---|---|---|---|---|---|---|---|---|---|---|---|---|---|---|
| ban date | Apr-23 | Apr-23 | Nov-22 | Apr-23 | Nov-22 | Nov-22 | Nov-22 | Apr-23 | Nov-22 | Nov-22 | Nov-22 | Nov-22 | Nov-22 | Nov-22 | Nov-22 | Nov-22 | Mar-23 |
| duration (days) | 822 | 822 | 273 | 822 | 973 | 973 | 426 | 822 | 457 | 457 | 426 | 973 | 457 | 457 | 973 | 92 | 31 |

Fig. 1. Global Bans on ChatGPT and Their Duration. Ban durations were calculated from each country's ChatGPT ban start date to either the official unban date or, if ongoing, to a fixed reference date (July 01, 2025). Durations are shown in calendar days.

Privacy concerns
OpenAI services/infrastructure
Country restrictions
Political concerns

### *B. AI and data protection laws in EU, UK, USA, China and GCC*

Key government-led milestones in AI across the EU, UK, USA, China and GCC demonstrate increasing national engagement with ethical, strategic, and sector-specific AI integration. As shown in Fig. 2, these developments include the establishment of AI authorities, national strategies, public-private partnerships, and regulatory developments. Data protection laws such as the GDPR (EU), UK Data Protection Act, and regional equivalents in the GCC (e.g., PDPL in Saudi Arabia and Bahrain) form the legal foundation for trustworthy AI implementation. [8]

Concurrently, several GCC countries—including the UAE, Saudi Arabia, and Qatar—have introduced national AI strategies that identify education as a priority sector. These are often supported by dedicated institutions like the UAE's Ministry of AI and Saudi Arabia's SDAIA, reflecting strong governmental commitment to AI-driven educational transformation. Many countries are in the process of drafting or refining regulatory frameworks, signaling global momentum toward more structured AI governance by 2025 [9].

The timeline reveals a clear trend of accelerating institutionalization and strategic commitment to AI across regions, particularly in the GCC countries, where rapid developments reflect a strong top-down push for AI-driven innovation. In contrast, the EU and UK emphasize regulatory structure and ethical alignment, suggesting a divergence in global AI governance approaches: GCC prioritizes implementation and growth, while EU/UK stress control and compliance—a dynamic that shapes how trustworthiness and adoption in education evolve regionally.

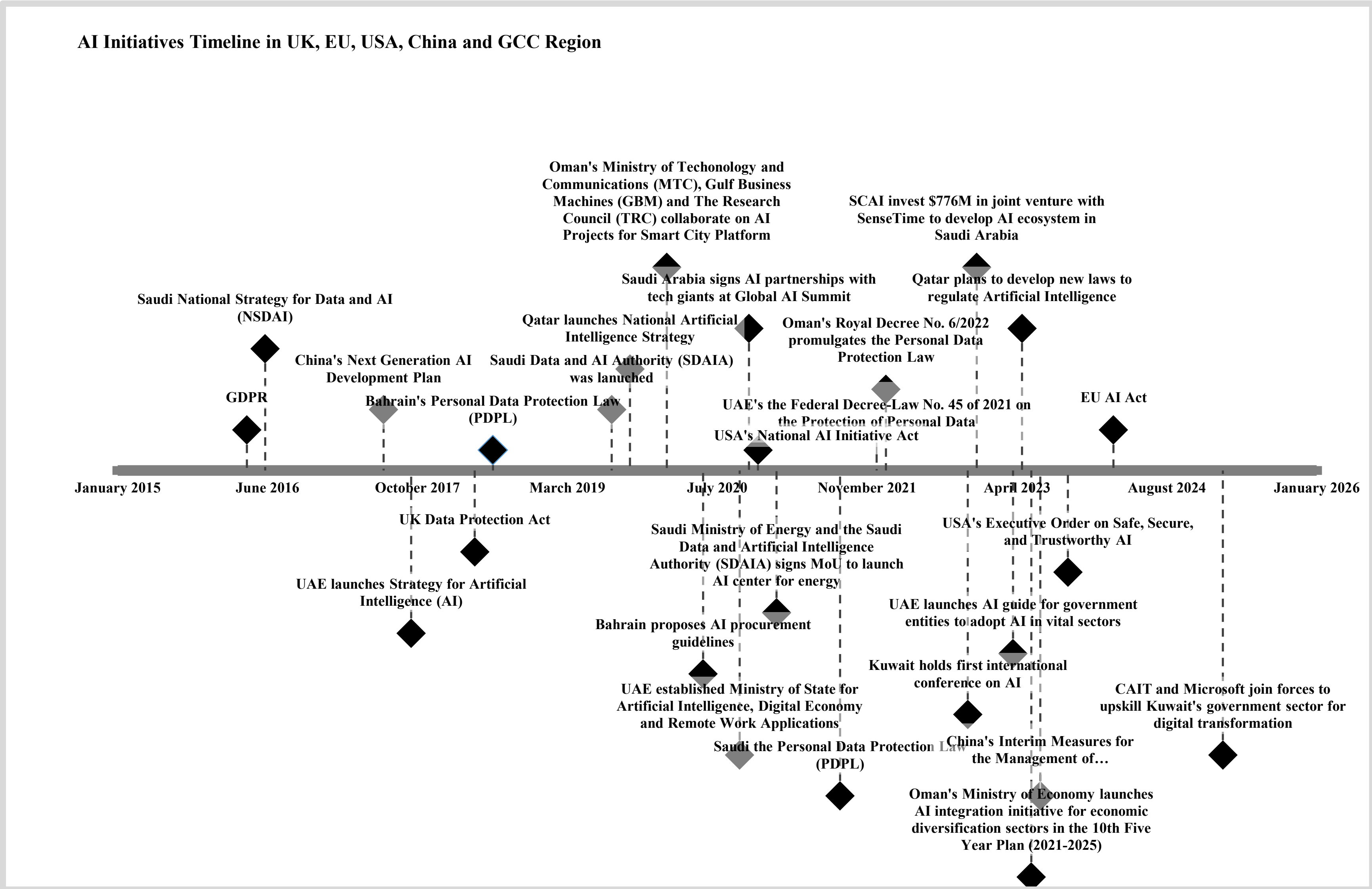

Fig. 2. Key legal frameworks and AI-related initiatives across EU, UK, USA, China and GCC countries.

## C. Alignment with International Standards and Ethical Principles

Legal and ethical standards are converging to guide responsible AI use in education. Scholars such as Li et al. advocate applying ethical principles from biomedical fields—beneficence, justice, autonomy, and non-maleficence—to LLM deployment [10]. Similarly, März et al. emphasize compliance with European legal frameworks, including the GDPR and the EU AI Act, which classifies AI systems by risk level and imposes documentation and transparency obligations [11].

TABLE I presents key ethical and operational principles structured across three levels: general values (e.g., human dignity, autonomy, fairness), user responsibilities (e.g., safe data handling, responsible usage), and organizational governance (e.g., promoting value and collaboration). These principles are grounded in guidance from European regulatory bodies like the High-Level Expert Group on AI (HLEG) and the European Group on Ethics (EGE) and are particularly relevant in sectors such as education and healthcare [12].

TABLE I. ETHICAL AND OPERATIONAL PRINCIPLES FOR THE TRUSTWORTHY USE OF LLMs [12]

| **Principle** | | **Responsible body** |
|---|---|---|
| ***General ethical considerations*** | - human dignity, autonomy, responsibility, justice, equity and responsibility, democracy, rule of law and accountability, security, safety, bodily and mental integrity, data protection and privacy, sustainability | The European Group on Ethics in science and new technologies (EGE) |
| | - autonomy, prevention of harm, fairness, and explicability<br>- accountability, human agency and<br>- oversight, technical robustness and safety, privacy and data governance, transparency, diversity, non-<br>- discrimination and fairness, societal and environmental wellbeing | The High-Level Expert Group (HLEG) |
| ***User principles*** | - Take appropriate measures to ensure safe input of data<br>- Continuously learn how to use LLMs effectively<br>- Know who to consult when facing concerns and report issues | - training networks and/or centres of expertise inside the European medicines regulatory network<br>- information security team and/or Data Protection Officer (DPO) |
| ***Organizational principles*** | - Define governance that helps users have a safe and responsible use<br>- Help users maximize value from LLMs<br>- Collaborate and share experiences | - European Specialized Expert Community (ESEC)<br>- EU Network training center |

## D. Regional Approaches to AI in Higher Education

The adoption of LLMs in education differs by region. In the GCC, countries like the UAE, Oman, and Saudi Arabia are rapidly advancing AI integration through personalized learning systems, AI-driven academic support, and the establishment of AI-focused institutions. The UAE's creation of the Mohamed bin Zayed University of Artificial Intelligence exemplifies this institutional investment [9].

In contrast, the EU and UK emphasize a cautious, ethics-oriented approach rooted in regulatory oversight. According to the Oxford Business Group, this distinction reflects differing priorities: while GCC states focus on leveraging AI for development and innovation, European jurisdictions prioritize compliance and risk management [13]. TABLE II compares these approaches, highlighting both common commitments—such as data protection and ethical alignment—and differences in regulatory maturity, institutional support, and implementation pace.

TABLE II. COMPARATIVE ANALYSIS OF AI REGULATIONS IN EDUCATION ACROSS SELECTED COUNTRIES

| | | |
|---|---|---|
| **Similarities:** | ***Focus on Data Protection*** | Almost all countries in Fig. 2, including the EU, UK, UAE, Saudi Arabia, Oman, and Qatar, have implemented data protection laws (e.g., GDPR, PDPL) as a foundational legal framework for AI deployment, including in education. |
| | ***Commitment to AI Strategy*** | Many countries (e.g., UAE, Saudi Arabia, Qatar, Oman) have national AI strategies that emphasize the integration of AI into education. These strategies often mention personalized learning, automation, and innovation in pedagogy. |
| | ***Alignment with Ethical Principles*** | The use of AI, particularly LLMs like ChatGPT, is expected to follow ethical guidelines such as fairness, transparency, and accountability—consistent with broader European and global discussions on AI ethics. |
| **Differences:** | ***Regulatory Maturity*** | The EU AI Act is currently the most comprehensive and binding framework, whereas other regions such as Bahrain, Oman, and Kuwait are still in the draft or planning phases for their national AI regulations. |
| | ***Institutional Support*** | Some countries (e.g., UAE with its dedicated AI ministry, and Saudi Arabia through SDAIA) have established specialized bodies to oversee AI adoption, while others rely on broader digital governance frameworks. |
| | ***Implementation Pace*** | GCC countries are rapidly implementing AI in education with strong government backing, whereas Western nations like the UK adopt a more cautious and consultative approach, emphasizing regulatory oversight and ethical reviews. |

## E. AI Compliance vs. Ethical Considerations

TABLE III distinguishes between legal compliance—focused on enforceable norms such as liability and sectoral regulations—and ethical frameworks, which emphasize aspirational values like fairness, accountability, and human rights. While compliance ensures institutions meet minimum legal requirements, ethics seeks to promote responsible and value-driven AI use that extends beyond what the law mandates.

TABLE III. LEGAL VS ETHICAL FRAMEWORKS IN AI GOVERNANCE

| | Aspects Related to Legal and Ethical Governance of AI | Laws / ethical frameworks |
|---|---|---|
| **AI Compliance as a Legal Framework** | - Legal risks<br>- Liability<br>- Enforcement mechanisms for AI developers and users | - Data protection laws (e.g., GDPR, CCPA)<br>- AI-specific regulations (e.g., EU AI Act, China's AI Guidelines)<br>- Sectoral compliance (e.g., AI in healthcare, finance, education)<br>- International governance frameworks (e.g., OECD AI Principles, UNESCO AI Ethics) |
| **AI Ethics as a Broader Philosophical and Societal Concern** | - Fairness, Accountability, and Transparency (FAT)<br>- Bias mitigation and non-discrimination<br>- Human rights and AI decision-making<br>- Moral responsibility in AI development | - IEEE Ethically Aligned Design<br>- UNESCO AI Ethics Guidelines<br>- Asilomar AI Principles |

In educational contexts, this distinction becomes critical. Legal compliance typically governs issues such as the protection of student data, algorithmic transparency in grading systems, and adherence to institutional privacy laws. Ethical considerations, however, concerns how AI is used to uphold academic integrity, prevent bias in assessment, and preserve the role of human judgment in learning environments.

Scholars remain divided on whether ethics should be codified into law or treated as a complementary framework [14], [15], [16]. What is increasingly recognized, however, is that both dimensions are essential: legal frameworks establish accountability and enforceability, while ethical considerations ensure human-centered design and pedagogical fairness

Nonetheless, there is growing consensus that ethical and legal approaches must work in tandem to foster trustworthy AI adoption in education. Effective governance will require interdisciplinary collaboration, harmonized standards, and ongoing research into AI's pedagogical and societal impacts.

## IV. TRUSTWORTHINESS IN AI LAWS

Building trust in AI requires more than technical excellence—it depends on governance systems that ensure ethical, transparent, and accountable use of the technology. Across jurisdictions, regulatory bodies are developing diverse strategies to embed trustworthiness into AI deployment, particularly in sensitive sectors like education. These strategies include requirements for explainability, fairness, human oversight, and data protection. TABLE IV provides a comparative overview of how selected countries and regions institutionalize trustworthiness through their regulatory bodies and policy frameworks, reflecting both shared principles and region-specific priorities.

TABLE IV. GLOBAL REGULATORY BODIES AND TRUSTWORTHINESS PRINCIPLES IN AI GOVERNANCE

| Country | Regulatory Body | Trustworthiness Aspects |
|---|---|---|
| **UK** [17], [18], [19] | - The UK government follows a **sectoral approach**, with **Ofqual** (for education) and the **ICO** (Information Commissioner's Office) ensuring compliance. | - Transparency<br>- Fairness & Non-Discrimination<br>- Accountability<br>- Safety & Robustness |
| **EU** [18], [20] | - The **European AI Office** will oversee compliance across EU member states. | - Risk-Based Approach<br>- Transparency & Explainability<br>- Human Oversight<br>- Ethical & Legal Compliance |
| **GCC** [21] | - **UAE:** Artificial Intelligence Office, Ministry of Education<br>- **Saudi Arabia:** Saudi Data & AI Authority (SDAIA)<br>- **Bahrain:** Ministry of Transportation and Telecommunications | - Ethical & Safe AI<br>- Data Security & Privacy<br>- AI Bias Prevention<br>- Human-Centric AI |
| **USA** [22], [23] | - National Institute of Standards and Technology (**NIST**) – AI Risk Management Framework (**AI RMF**)<br>- White House Executive Order on AI Safety<br>- Federal Trade Commission (**FTC**) – AI Fairness & Transparency Regulations<br>- Food and Drug Administration (**FDA**) – AI in Healthcare Compliance | - AI fairness, accountability, and transparency<br>- Risk management & ethical AI<br>- Data privacy & security<br>- Explainability & bias mitigation |
| **China** [24], [25] | - Cyberspace Administration of China (**CAC**) – AI Algorithm Regulations<br>- National AI Standardization Group<br>- Ministry of Science and Technology (**MOST**) – AI Innovation & Ethics Policy<br>- Chinese Academy of Sciences (**CAS**) – AI Ethics Guidelines | - Strict AI content control & regulation of algorithms<br>- AI reliability & data protection<br>- Social impact assessment of AI technologies<br>- AI safety |
| **Asia (General)** [26], [27], [28] | - **Singapore:** AI Governance Framework, Personal Data Protection Commission **(PDPC)**<br>- **Japan:** AI Ethics Guidelines (Ministry of Economy, Trade, and Industry (METI))<br>- **South Korea:** AI Policy & Governance (National Information Society Agency (NIA) & Ministry of Science and ICT (MSIT)) | - Responsible AI use & transparency<br>- Data protection<br>- Ethical AI innovation & safety<br>- AI reliability & public trust |

This comparative analysis shown in TABLE IV, reveals a growing global consensus around key trustworthiness principles—such as transparency, accountability, and human oversight—yet the implementation methods vary by region. For instance, the EU adopts a structured, risk-based regulatory model with legal enforceability, while the UK emphasizes sector-specific oversight. In contrast, GCC countries integrate cultural and ethical values into AI governance, and China enforces strict algorithmic control tied to national security. Meanwhile, the USA leans on institutional coordination and voluntary frameworks like NIST's AI RMF [23]. These differences underscore the importance of context-sensitive approaches to regulating AI in education, especially when aligning international cooperation with domestic priorities. As LLMs continue to shape learning environments, understanding these regulatory divergences is essential to developing trustworthy and globally relevant AI policies.

TABLE V illustrates how key trustworthiness principles—transparency, fairness and non-discrimination, reliability and safety, accountability, data privacy, and human oversight—are emphasized across AI regulatory frameworks in selected global regions. It provides a comparative view of each country or region's approach to building trustworthy AI, particularly relevant to education and LLMs.

TABLE V. MAPPING TRUSTWORTHINESS ASPECTS OF AI REGULATION ACROSS REGIONS

| **Trustworthiness Aspect** / **Country** | **Transparency** | **Fairness and Non-Discrimination** | **Reliability and Safety** | **Accountability** | **Data Privacy** | **Human Oversight** |
|---|---|---|---|---|---|---|
| **UK** [17], [18], [19] | ✓ | ✓ | ✓ | ✓ | ✓ | ✓ |
| **EU** [18], [20] | ✓ | ✓ | ✓ | ✓ | ✓ | ✓ |
| **GCC** [21] | | ✓ | ✓ | | ✓ | ✓ |
| **USA** [22], [23] | ✓ | ✓ | ✓ | ✓ | ✓ | |
| **China** [24], [25] | | | ✓ | | ✓ | |
| **Asia (General)** [26], [27], [28] | ✓ | | ✓ | | ✓ | |

**✓ indicates the aspect is explicitly addressed or emphasized in regulatory or ethical AI frameworks. Blank cells reflect limited or emerging regulatory clarity in that area.**

Based on the analysis of TABLE V, there is growing international alignment around key trustworthiness principles in AI governance—especially reliability, safety and data privacy—but notable differences persist in emphasis and regulatory maturity across regions. The UK and EU demonstrate the most comprehensive frameworks, addressing all six trustworthiness aspects. The USA also covers a broad spectrum, focusing strongly on transparency, fairness, accountability, and data privacy, but shows less emphasis on human oversight compared to its European counterparts. GCC and Asian regions reflect emerging efforts, with selective focus areas such as ethical AI use and cultural values, while China prioritizes content control, safety, and national security, showing limited alignment with broader ethical frameworks. These variations underscore the ongoing need for interoperable and globally coherent AI standards, especially in sensitive domains like education.

## V. FUTURE COMPLIANCE PATHWAYS: A FRAMEWORK FOR AI GOVERNANCE IN GCC EDUCATION

With the rapid adoption of AI in education across GCC countries, there is an urgent need for governance models that are structured, legally aligned, and sensitive to local cultural contexts. To address this, we propose a Compliance-Centered AI Governance Framework for the GCC region. The framework includes a tiered typology reflecting maturity levels of AI alignment with legal, ethical, and cultural standards, along with an actionable checklist for educators, regulators, and developers. It integrates global trust principles (e.g., fairness, transparency, accountability) with local legal frameworks, religious considerations, and societal values to ensure safe and lawful AI use in education. Fig. 3 outlines five ascending tiers of compliance, from basic legal alignment to full legal-ethical integration.

- **Regulators** (e.g., Bahrain's Telecom Ministry, UAE's AI Ministry, SDAIA) can **adopt this typology** to assess AI tools submitted for educational approval.
- **Developers** working in the GCC can align their products with **Tier 5 criteria** for improved policy compliance and regional market fit.
- **Institutions** can use the following checklist shown in TABLE VI as a self-assessment tool to prepare for internal audits, procurement reviews, or ministry approvals.

The proposed framework not only addresses immediate compliance needs within individual GCC states but also lays the groundwork for broader regional harmonization. As AI continues to reshape education globally, aligning national strategies with international benchmarks—such as the UNESCO Recommendation on the Ethics of AI and the OECD AI Principles—can enhance interoperability, build public trust, and position GCC countries as active contributors to global AI governance [29]. Emerging tools like regulatory sandboxes—already piloted in countries like the UAE—offer a controlled environment for safely testing AI in education while remaining legally compliant [30]. Future efforts should focus on fostering collaborative policymaking, capacity-building in AI literacy, and shared regulatory tools across the region to ensure consistent, ethical, and forward-looking use of AI in education.

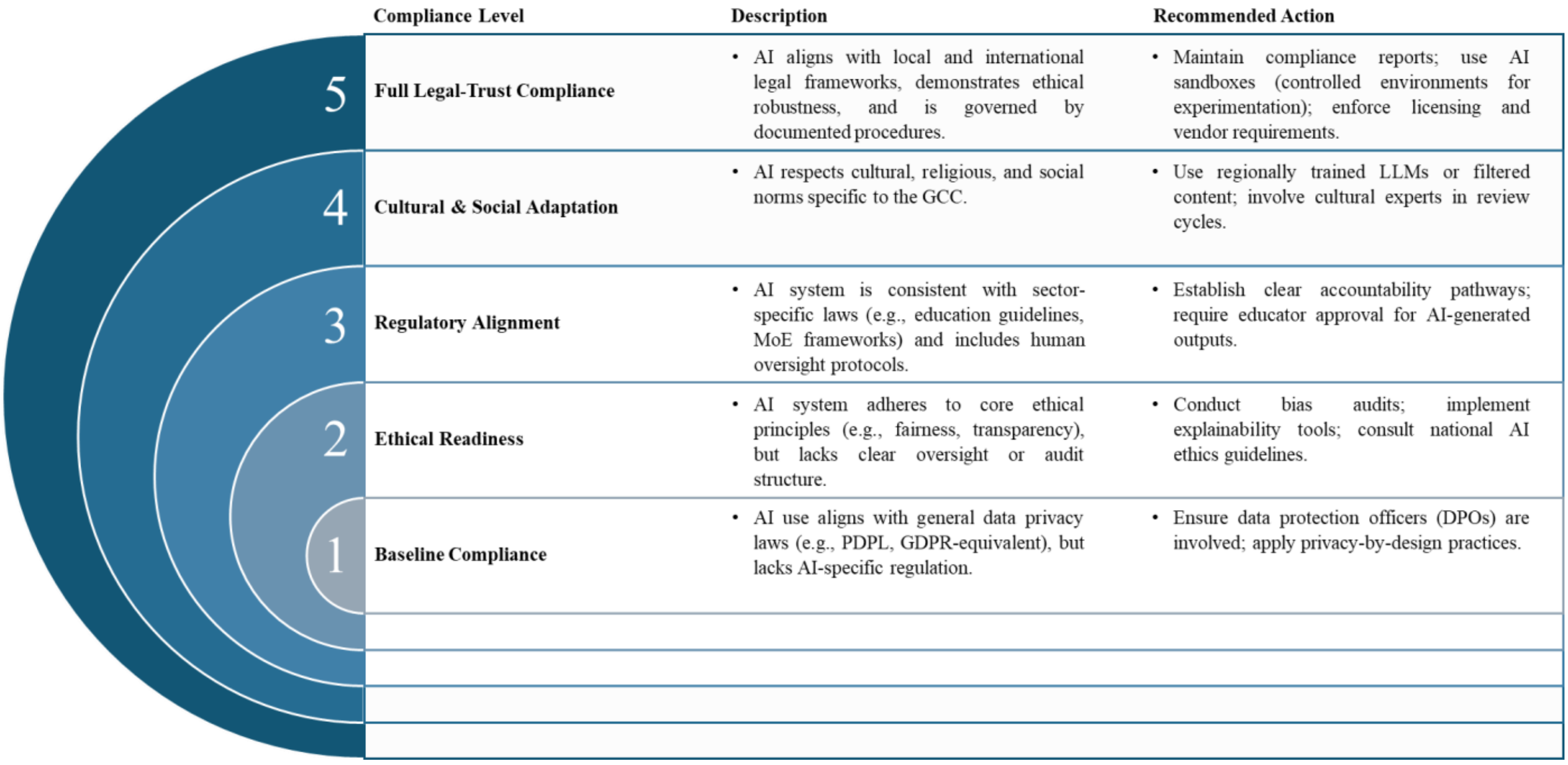

Fig. 3. Proposed GCC AI Education Compliance Framework: Aligning AI Use with Legal and Ethical Compliance

TABLE VI. AI READINESS & COMPLIANCE CHECKLIST FOR GCC EDUCATION INSTITUTIONS

| | Checklist Item | Why It Matters |
|---|---|---|
| ☐ | Data protection assessment conducted (PDPL / GDPR / consent) | Ensures lawful collection, storage, and sharing of student data. |
| ☐ | Bias and fairness evaluation done | Prevents discriminatory outputs and upholds fairness in assessments. |
| ☐ | Transparent explanation of AI decision-making available to educators and students | Builds trust and enables user understanding. |
| ☐ | Human oversight policy in place | Ensures educators remain accountable for teaching and grading. |
| ☐ | Content culturally reviewed and aligned with local values | Avoids reputational risk and ensures contextual appropriateness. |
| ☐ | AI vendor/technology audited and documented | Creates accountability and aligns with procurement regulations. |
| ☐ | Training provided to educators on AI use and risks | Builds internal capacity for safe and effective implementation. |

## VI. CONCLUSION

The rapid integration of LLMs in education presents both opportunities for innovation and significant legal, ethical, and cultural challenges—particularly in regions like the GCC where AI governance is still maturing. This paper has provided a comparative analysis of AI-related regulatory frameworks across global regions, with a special focus on how trustworthiness principles—such as transparency, fairness, accountability, and human oversight—are operationalized in education law.

While the European Union leads with a structured, risk-based regulatory approach, other regions like the GCC are advancing through national AI strategies, emerging data protection laws, and institution-specific initiatives. Despite these regional differences, a shared global emphasis on ethical alignment and legal compliance is beginning to take shape.

To support this momentum, we introduced a Compliance-Centered AI Governance Framework tailored to the GCC context. This forward-looking model offers a tiered typology and checklist to help regulators, developers, and institutions align AI tools with both international standards and local expectations. Such tools are essential for ensuring that AI adoption in education remains responsible, inclusive, and trustworthy.

Ultimately, effective AI governance in education must go beyond abstract principles or top-down mandates. It requires context-aware implementation, sustained human oversight, and interdisciplinary collaboration across legal, technical, and pedagogical domains. As LLMs continue to reshape learning environments, countries that invest in adaptive, value-driven governance frameworks will be better positioned to harness AI's potential while safeguarding public trust.